\documentclass[12pt]{article}

\usepackage{geometry}
\usepackage[font=footnotesize,margin=1.5cm]{caption}

\usepackage{amsmath}
\usepackage{amssymb}
\usepackage{amsthm}
\usepackage{graphicx}

\title{A practical heuristic for finding graph minors}

\author{Jun Cai, Bill Macready, Aidan Roy \\
D-Wave Systems Inc.}
\date{\today}

\begin{document}
\maketitle

\abstract{We present a heuristic algorithm for finding a graph $H$ as a minor of a graph $G$ that is practical for sparse $G$ and $H$ with hundreds of vertices. We also explain the practical importance of finding graph minors in mapping quadratic pseudo-boolean optimization problems onto an adiabatic quantum annealer.}


\section{Introduction}
 
Graph minors are an extremely important theoretical topic due to the work of Robertson and Seymour \cite{RS95}, which implies that when $H$ is fixed, there is a polynomial-time algorithm for determining whether or not $G$ contains $H$ as a minor. Many important applications have been shown to be polynomial-time or at least fixed-parameter tractable as a result of the graph minors project. However, when both $H$ and $G$ are part of the input, the minor-embedding problem is NP-hard. Currently the best known algorithm has running time $O(2^{(2k+1)\log k} |H|^{2k} 2^{2|H|^2} |H|)$, where $k$ is the branchwidth of $G$ \cite{ADFST11}. Moreover, algorithms coming from the Robertson-Seymour theory are non-constructive for general $H$, and the constant factors used in the algorithms are prohibitively large. 

As a result, none of the known exact algorithms for minor-embedding are practical for more than tens of vertices. In this paper we focus instead on heuristic techniques: we find a minor with some probability, without attempting an exhaustive search and without attempting to prove minor-exclusion in the case of failure. 

It seems that heuristic minor-embedding for arbitrary $H$ has received little attention in the graph theory literature, possibly due to a lack of known applications. (See \cite{KR96} for one algorithm when $G$ is an expander graph.) In fact, minor-embedding is central to mapping optimization problems to the adiabatic quantum machine developed by D-Wave Systems. More precisely, a quadratic boolean optimization problem can be mapped onto the D-Wave hardware if and only if the graph of variable interactions in the optimization problem is a minor of the graph of qubit interactions in the D-Wave hardware. We explain the details of this application in Section 2.

The success of a heuristic minor-embedding algorithm depends heavily on that fact that, if $H$ is significantly smaller than $G$, then there are probably a large number of distinct $H$-minors in $G$. As a simple example, consider minor-embedding $P_n$ (the path-graph on $n$ vertices) in $P_{2n}$. Up to reversing the order of the path, each vertex $x_i$ of $P_n$ is represented by a subpath starting at a vertex $y_i$ of $P_{2n}$. Therefore the number of minor-embeddings is twice the number of integer sequences $1 \leq y_1 < y_2 < \ldots < y_n < y_{n+1} \leq 2n+1$, which is $2\binom{2n+1}{n+1} = \Omega(4^n/\sqrt{n})$. That is, the number of minor-embeddings grows exponentially in $n$. The general question of the number of $H$-minors in $G$ does not appear to have been studied, but probabilistic arguments have been used to show the existence of large clique minors in various graphs \cite{Fountoulakis2008,Fountoulakis2009,Krivelevich2009}.

This paper is organized a follows. In Section 2, we explain the application of minor-embedding to D-Wave's adiabatic quantum annealer. In Section 3, we present the heuristic embedding algorithm. In Section 4, we present some performance results, in which the algorithm has been used to find minors when $G$ and $H$ have up to $500$ and $200$ vertices respectively. In Section 5, we present a modified version of the algorithm more suited to very large graphs. Future work is discussed in Section 6.

\section{Minor-embedding and quantum annealing}

Our motivation for studying heuristic minor-embedding stems from its importance in mapping optimization problems to D-Wave's quantum computer. More details can be found in \cite{Choi2008,Bian2010}.

The D-Wave Two device solves problems in the form on an \emph{Ising model}: given parameters $h \in \mathbb{R}^n$ and $J \in \mathbb{R}^{n \times n}$, minimize the energy function
\[
E(z) = h^Tz + z^TJz
\]
subject to $z \in \{\pm 1\}^n$. This is a quadratic optimization over the boolean variables (the \emph{qubits}) with real coefficients, an NP-complete problem \cite{Barahona1982,Boros2002}. Typically we use the D-Wave machine to solve optimization problems by first expressing the problem using boolean variables, then reducing to quadratic form, and finally mapping to the D-Wave hardware. 

D-Wave Two solves these problems using adiabatic quantum annealing \cite{Farhi2000}: a quantum system is slowly evolved until the ground state of the system encodes the solution to the optimization problem \cite{Harris2010,Johnson2011}. Adiabatic quantum computation is polynomially equivalent to gate-model quantum computation \cite{Aharonov2007}, although the restriction to Ising models means that D-Wave machine is not universal. Experiments show the D-Wave machine at its current scale is competitive with the best classical algorithms for solving Ising models \cite{McGeoch2013,Ronnow2014,Google2014,Selby2013}.
 
\begin{figure}[ht]
\begin{center}
\includegraphics[scale = 0.3]{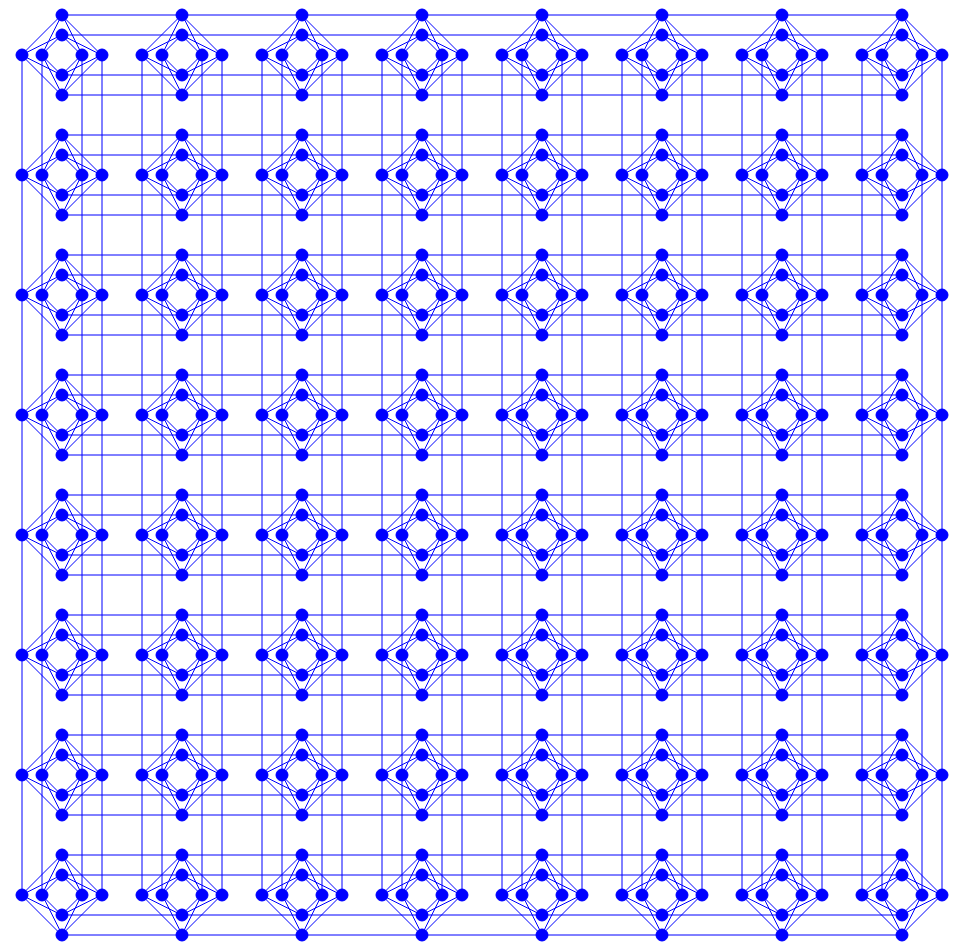}
\end{center}
\caption{``Chimera": the graph of qubit interactions in D-Wave's machine.}
\label{fig:chimera}
\end{figure}

Not every Ising model can be directly applied on the D-Wave machine: the interactions available between variables are restricted by engineering limitations. Define the graph of an Ising model $[h,J]$ as follows: the vertices are the variables $z_1,\ldots,z_n$, with an edge between $z_i$ and $z_j$ if $J_{ij}$ is nonzero. The current D-Wave Two hardware graph (Figure \ref{fig:chimera}) has up to 512 qubits, with each qubit adjacent to at most $6$ others.

In general, an Ising model we would like to solve will not have this particular graph structure. In order to map to the hardware graph, we use multiple qubits in the hardware graph ({\em physical qubits}) to represent the same variable in the problem Ising model ({\em logical qubits}). 

Suppose we want physical qubits $z_1$ and $z_2$ to represent the same logical qubit. By setting $J_{12} = -\infty$ in the hardware Ising model, we ensure that $q_1$ and $q_2$ take the same value at the optimal solution.\footnote{In practice, $J_{ij}$ values are limited to the range $[-1,1]$, so we set $J_{12} = -1$ and scale other interactions down to smaller absolute values.} In this way, a logical qubit can be modelled by any collection of physical qubits provided those physical qubits form a connected subgraph of the hardware graph. A non-zero interaction between two logical qubits can be represented precisely when there is an edge between the corresponding connected subgraphs of physical qubits. In summary: an Ising model $[h,J]$ can be represented on the hardware if and only if the graph of $[h,J]$ is a minor of the hardware graph.

In practice, the hardware graph $G$ produced by D-Wave is slightly different in each processor as certain qubits of insufficient quality are disabled. Moreover future versions of the hardware may have entirely different graphs. So, we are interested in algorithms in which both $G$ and $H$ are part of the input. That being said, the structure of the Chimera graph makes it more suited to certain algorithms. Firstly, the hardware graph is designed to have large treewidth \cite{Bunyk2014}, meaning that exact algorithms are of no practical use at the current scales. Secondly, the Chimera graph has a very large automorphism group (size $4!^{16} \cdot 8$), which in some sense reduces the number of choices in a heuristic algorithm. And thirdly, the graph is sparse, meaning that shortest paths can be computed in linear time.

Also of practical importance is the fact that not all minor-embeddings result in the same hardware performance. In general for a given minor-embedding problem we would like to minimize either the maximum number physical qubits representing any logical qubit, or minimize the total number of physical qubits used.

\section{A minor-embedding algorithm}

A minor-embedding of $H$ in $G$ is defined by a function $\phi:V(H) \rightarrow 2^{V(G)}$ (called a {\em model}) such that 
\begin{enumerate}
\item for each $x \in V(H)$, the subgraph induced by $\phi(x)$ in $G$ is connected; 
\item $\phi(x)$ and $\phi(y)$ are disjoint for all $x \neq y$ in $V(H)$; and 
\item if $x$ and $y$ are adjacent in $H$, then there is at least one edge between $\phi(x)$ and $\phi(y)$ in $G$. 
\end{enumerate}
We call $\phi(x)$ the {\em vertex-model} of $x$ and say that $\phi(x)$ {\em represents} $x$ in $G$.

Our heuristic algorithm for minor-embedding proceeds by iteratively constructing a vertex-model for each vertex of $H$, based on the locations of its neighbours' vertex-models. We begin by describing the method for finding a good vertex-model. 

\begin{figure}[ht]
\centerline{\includegraphics[scale = 0.4]{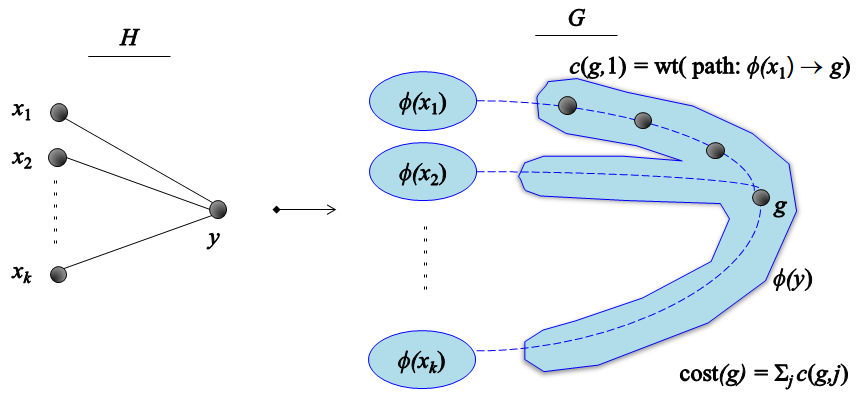}}
\caption{Evaluating the potential cost of each vertex $g \in V(G)$ as the root of $\phi(y)$. The vertex of smallest cost is selected as the root, and the new vertex-model is defined to be the union of the shortest paths from the root to the neighbour vertex-models.}
\label{fig:vertadd}
\end{figure}

Suppose we want a vertex-model $\phi(y)$ for $y \in V(H)$, and $y$ is adjacent to $x_1,\ldots,x_k$ which already have vertex-models $\phi(x_1),\ldots,\phi(x_k)$. A ``good'' vertex-model might be one that ensures that $\phi(y)$ shares an edge with each of $\phi(x_1),\ldots,\phi(x_k)$, while minimizing the number of vertices in $\phi(y)$. Doing so leaves as much room as possible for other vertex-models yet to be determined. To this end, for each $x_j$, compute the shortest-path distance from $\phi(x_j)$ to every vertex $g$ in the subgraph of $G$ of ununsed vertices (i.e. the vertices not in any vertex-model). Record this information as a cost $c(g,j)$. Then, select the vertex $g^*$ that has the smallest total sum of distances $\sum_j c(g,j)$, and declare $g^*$ to be the {\em root} of the new vertex-model $\phi(y)$. Finally, identify a shortest path from $g^*$ to each $\phi(x_j)$, and take the union of those paths as $\phi(y)$. This method is depicted in Figure \ref{fig:vertadd}.

It will often be the case that no such $g^*$ exists (i.e. no vertex in $G$ has a path to each of $\phi(x_1),\ldots,\phi(x_n)$ using only unused vertices). To circumvent this problem, we temporarily allow vertex-models to overlap: we allow multiple vertices of $H$ to be represented at the same vertex of $G$. A ``good" vertex-model is then defined to be one that uses a minimal number of vertices at which multiple vertex-models appear. To compute these costs, we use a {\em weighted} shorted paths calculation, in which the weight of a vertex in $G$ grows exponentially with the number of vertices of $H$ represented there. Let $D$ be the diameter of $G$. For each $g \in V(G)$, define a vertex weight
\[
\mathrm{wt}(g) := D^{|\{i: g \in \phi(x_i)\}|}.
\]
Then define the weight of a path to be the sum of the weights of the vertices in that path, and define the cost $c(g,j)$ to be the smallest weight of a path from $\phi(x_j)$ to $g$. (Note: unless $g$ is in $\phi(x_j)$, we exclude the weight of $\phi(x_j)$ from $c(g,j)$, since no vertex of $\phi(x_j)$ will be added to the new vertex-model $\phi(y)$.) Computing shortest paths in this way ensures that the root of $\phi(y)$ is chosen so that as few vertex-models overlap as possible.

The minor-embedding algorithm now proceeds as follows. In the initial stage, we choose a vertex order at random, and for each vertex $x$ we find a vertex-model based on the weighted shortest path distances to its neighbours as described above. If none of $x$'s neighbours have vertex-models yet, we choose a vertex of $G$ at random to be the vertex-model for $x$. 

After the initial stage, we try to refine the vertex-models so that no vertex of $G$ represents more than one vertex of $H$. We do this by iteratively going through the vertices of $H$ in order, removing a vertex-model from the embedding and reinserting a better one. In doing so, we select shortest paths such that the overlap between vertex-models typically decreases.

Once we have gone through all the vertices, we check to see if an improvement has been made, where improvement is measured by
\begin{itemize}
\item the largest number of vertices of $H$ represented at any vertex of $G$;
\item failing that, the total sum of the vertex-model sizes.
\end{itemize}
We stop when at most one vertex-model is represented at any vertex of $G$ (meaning that we have found a minor-embedding) or no improvement has been made after fixed number of iterations.

The algorithm is summarized in Figure \ref{fig:decompalg}.

\begin{figure}[p]
\begin{center}
\begin{minipage}{8cm}
\hrule
\footnotesize
\begin{tabbing}
\\
\underline{\textbf{findMinorEmbedding($G$,$H$)}} \\
\textbf{Input}: graph $H$ with vertices $\{x_1,\ldots,x_n\}$, graph $G$ \\
\textbf{Output}: vertex-models $\phi(x_1),\ldots,\phi(x_n)$ of an $H$-minor in $G$, or ``failure". \\
\\
randomize the vertex order $x_1,\ldots,x_n$ \\
set $stage$ := 1 \\
for $i$\= ~$\in \{1,\ldots,n\}$ do \\
	\> set $\phi(x_i)$ := $\{\}$ \\
while $\max_{g \in V(G)} |\{i: g \in \phi(x_i)\}|$ or $\sum_i |\phi(x_i)|$ is improving, or $stage \leq 2$ \\
	\> for $i$\= ~$\in \{1,\ldots,n\}$ do \\
		\>\> for $g$\= ~$\in V(G)$ do \\
		\>\>\> set $w(g) := \mathrm{diam}(G)^{|\{j \neq i: g \in \phi(x_j)\}|}$ \\
		\>\>$\phi(x_i)$ := findMinimalVertexModel($G,w,\{\phi(x_j): x_j \sim x_i\}$) \\
	\> set $stage := stage + 1$ \\
if $|\{i: g \in \phi(x_i)\}| \leq 1$ for all $g \in V(G)$ \\
\> return $\phi(x_1),\ldots,\phi(x_n)$ \\
else \\
\> return ``failure"
\\			
\\
\\
\underline{\textbf{findMinimalVertexModel($G, w, \{\phi(x_j)\}$)}} \\
\textbf{Input}: graph $G$ with vertex weights $w$, neighbouring vertex-models $\{\phi(x_j)\}$ \\
\textbf{Output}: vertex-model $\phi(y)$ in $G$ such that there is an edge between $\phi(y)$ and each $\phi(x_j)$  \\
\\
if all $\phi(x_j)$ are empty \\
	\> return random $\{g^*\}$ \\
for all $g \in V(G)$ and all $j$ do \\
	\> if $\phi(x_j)$ is empty \\
	\>\> set $c(g,j)$ := 0 \\
	\> else if $g \in \phi(x_j)$ \\
	\>\> set $c(c,g)$ := $w(g)$ \\
	\> else \\
	\>\> set $c(g,j)$ := $w$-weighted shortest-path distance($g,\phi(x_j)$) excluding $w(\phi(x_j))$ \\
set $g^* := \mathrm{argmin}_g \sum_j c(g,j)$ \\
return $\{g^*\} \cup \{$paths from $g^*$ to each $\phi(x_j)\}$ \\
\\
\end{tabbing}
\hrule
\end{minipage}

\caption{The heuristic for finding an $H$-minor in $G$. The notation $x_j \sim x_i$ indicates that $x_j$ and $x_i$ are adjacent in $H$.}
\label{fig:decompalg}
\end{center}
\end{figure}

\subsection*{Implementation and improvements}

There are several heuristic choices in implementation which are critical to improving the algorithm's performance. We describe those choices here.

\begin{figure}[ht]
\begin{center}
\includegraphics[scale = 0.4]{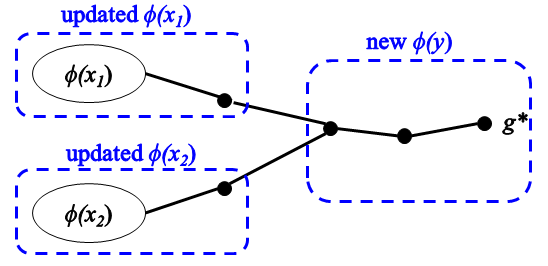}
\end{center}
\caption{Determining which vertices on a path from $\phi(x_i)$ to $g^*$ are added to which vertex model. Vertices appearing in multiple paths are added to $\phi(y)$, others are added to $\phi(x_i)$.}
\label{fig:vertex-model}
\end{figure}

\begin{itemize} 
\item {\bf Vertex-weighted shortest paths}:
We can compute the shortest path from a fixed vertex of $G$ to every other vertex using Dijkstra's algorithm. Dijkstra's algorithm as normally written does not apply to graphs with weighted vertices, but we can translate vertex weights into edge weights by considering a directed graph in which the weight of every arc is the weight of its head. 

\item {\bf Shortest paths to subsets of vertices}: 
Note that when we compute the distance from a vertex $g$ to a vertex-model $\phi(x_i)$, it is not necessary to compute the distance to each vertex within the vertex-model individually. Instead, we add a dummy vertex $\tilde{x_i}$ adjacent to every vertex in $\phi(x_i)$, and compute the shortest-paths distances to $\tilde{x_i}$.

\item {\bf Choice of vertex-model on paths}: Note that after selecting a vertex $g^*$ and shortest paths for a vertex-model $\phi(y)$ to represent vertex $y$, the vertices on the path from $g^*$ to a neighbour vertex-model $\phi(x_j)$ could in fact be added to $\phi(x_j)$ rather than $\phi(y)$. One heuristic for choosing which vertex-model to add vertices of $G$ to is the following: if $v \in V(G)$ appears in more than one shortest path, say the paths to $\phi(x_j)$ and $\phi(x_k)$, then add $v$ is added to $\phi(y)$. (This is preferable to adding $v$ to both $\phi(x_j)$ and $\phi(x_k)$, since fewer vertices of $H$ are represented at $v$.) On the other hand, if $v$ appears in only a single shortest path, say the path to $\phi(x_j)$, then add $v$ to $\phi(x_j)$ (as doing so may allow for shorter paths to other neighbours of $x_j$.) This heuristic is illustrated in Figure \ref{fig:vertex-model}.

\item {\bf Random root selection}:
Instead of choosing a root $g^*$ for a vertex-model to be the vertex with minimal cost, we may choose $g^*$ randomly, with the probability of choosing $g$ proportional to $e^{-cost(g)}$. This randomness helps avoid getting stuck in local optima.
\end{itemize}

\section{Running time and performance}

The largest part of the algorithm's running time is computing the shortest-path distances between vertex model and other vertices of $G$. Note that the shortest paths must be recomputed in every iteration, as the weights of the vertices change.

Assume $H$ has $n_H$ vertices and $e_H$ edges, and $G$ has $n_G$ vertices and $e_G$ edges. Dijkstra's algorithm has run time $O(e_G + n_G \log n_G)$. Each iteration of the embedding algorithm calls Dijkstra $2e_H$ times (once for each ordered pair of adjacent vertices in $H$). We terminate if there is no improvement after a constant number of iterations, and the number of possible improvements before an embedding is found is at most $n_Hn_G$ (the worst possible sum of vertex-model sizes). Therefore the run time of the algorithm is $O(n_Hn_Ge_H(e_G + n_G \log n_G))$. In practice, the number of iterations of the algorithm before termination or success is very instance-dependent but typically much less than $n_Hn_G$. 

\begin{figure}[ht]

\begin{center}
\setlength{\unitlength}{1cm}
\begin{picture}(14.5,12)
\put(0.25,6){\includegraphics[scale = 0.35]{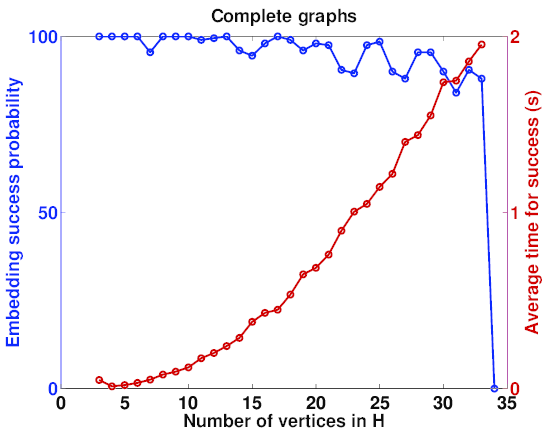}}
\put(0,11){a)}
\put(7.75,6){\includegraphics[scale = 0.35]{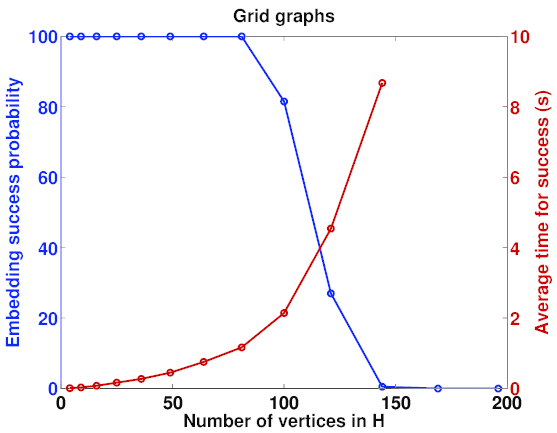}}
\put(7.5,11){b)}
\put(4.5,0){\includegraphics[scale = 0.35]{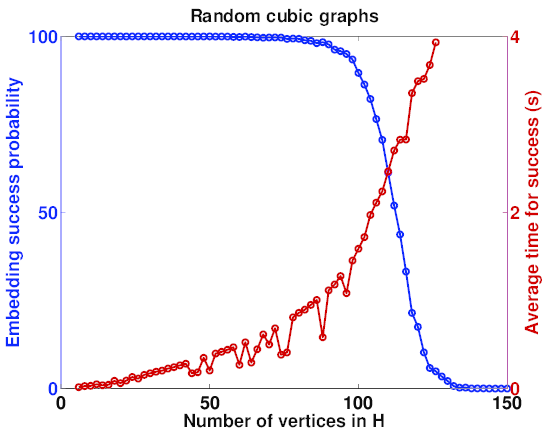}}
\put(4.25,5){c)}
\end{picture}
\end{center}
\caption{Performance of the heuristic minor-embedding algorithm in finding minors in the $512$-vertex Chimera graph $G$. The run time is in seconds using a single core of an Intel Xeon E5-2670 processor. (a) Complete graphs: using treewidth arguments it can be proved that $K_{33}$ is the largest complete graph that is a minor of $G$. This graph embeds with high probability in under two seconds. (b) Grid graphs: the $16\times 16$ grid graph ($256$ vertices) is the largest grid graph known to be a minor of $G$. (c) Random cubic graphs: each point in the plot represents the median case of $20$ graphs of that size.}
\label{fig:performance}
\end{figure}

To show the performance of the algorithm, we considered minor-embedding three different types of graph into the Chimera graph in Figure \ref{fig:chimera}: complete graphs, grid graphs, and random cubic graphs. For each type, we generated graphs in a range of sizes and ran the algorithm 100 times on each graph, recording the running time and whether or not a minor was found. The run times and success rates are given in Figure \ref{fig:performance}.

\section{Localized version}

The running time of the minor-embedding algorithm in Section 3 is dominated by the need to compute the length of the shortest path to every vertex of $G$, from every vertex-model, at every iteration. This is wasteful, as only one vertex will be selected as the root of a new vertex-model, and in many cases that choice will not change from the previous iteration. In this section we describe a  modification to the algorithm which typically searches for the root of a vertex-model locally rather than globally. This modification has two components.

First, we replace the series of shortest-path computations, one for each vertex-model, by a single computation in which the shortest paths from each vertex-model are computed simultaneously. We call this {\em multisource Dijkstra}. Recall that Dijkstra's algorithm grows a shortest-paths tree $T$ from a source $s$ by selecting the vertex $v$ closest to $s$ not in $T$, and adding $v$ to $T$. In multisource Dijkstra, we again grow a shortest-path trees $T_1,\ldots,T_k$ from each of our sources $s_1,\ldots,s_k$. Let $v_i$ be the vertex closest to $s_i$ and not in $T_i$, with distance $d_i$ to $s_i$. At each step of the algorithm, we select the index $i^*$ such that the distance $d_{i^*}$ is minimal among all $\{d_1,\ldots,d_k\}$. We add $v_{i^*}$ to $T_{i^*}$, update the distances to $s_{i^*}$, and repeat. At each vertex, we record whether or not it has been reached by each source, and we terminate when some vertex $v^*$ has been reached by all sources. Vertex $v^*$ will be the vertex whose maximal shortest-path distance is minimized. Note that in doing multisource Dijkstra, we no longer compute the shortest path to every vertex from every source: the algorithm is ``localized'' to a subset of the entire graph.

Second, we replace Dijkstra's algorithm in the shortest-path computation with an A* search algorithm \cite{Hart1968}. The target of the A* search for vertex $x_i$ is the root $g^*$ of the vertex-model $\phi(x_i)$ found in the previous iteration. (In the first iteration, there is no target.) The heuristic estimate of the distance from a vertex $g$ to the target $g^*$ is the unweighted shortest-path distance between $g$ and $g^*$ in $G$, which can be precomputed. We call this a {\em multisource} A* algorithm. A summary is given in Figure \ref{fig:multiastar}.

The practical effect of using multisource A* search in the minor-embedding algorithm is as follows: as before, when finding shortest paths, we first select vertices in $G$ that represent fewer vertices in $H$. However, when choosing between two vertices in $G$ that represent the same number of vertices in $H$, we chose the one that is closer to the previous root of the vertex-model. So, in situations where we do {\em not} improve the vertex-model in terms of the number of represented vertices, we find the root of the previous vertex-model very quickly. On the other hand if an improvement to the vertex-model {\em is} possible, we will find it in roughly the same time as we would using Dijkstra's algorithm.

\begin{figure}[p]
\begin{center}
\begin{minipage}{5.5in}
\hrule
\footnotesize
\underline{\textbf{multisource A*}} \\
\textbf{Input}: graph $\tt{G}$, edge weights $\{\tt{w(u,v)}\}_{uv \in E(G)}$, heuristic costs $\{\tt{h(v)}\}_{v \in V(G)}$, sources $\{\tt s(1),\ldots,s(k)\} \subseteq V(G)$ \\
\textbf{Output}: vertex $\tt{cv}$ such that $\max_{i=1}^k \tt{d(cv,s(i))}$ is minimal among all vertices \\

\begin{verbatim}
for each v in V(G), each source i:                            
    d(v,i) := infinity;                    // best known distance from v to i 
    est(v,i) := infinity;                  // heuristic distance 
    reached(v,i) := false;                 // node v reached from source i?

for each v in V(G):								
    min_est(v) := infinity;                // min. dist. to v among all i             

for each source i:
    d(s(i),i) = 0;
    min_est(s(i)) := 0;
    min_src(s(i)) := i;                    // index of source for min_est

while true:                                
    cv := argmin_{v in V(G)} min_est(v);   // current node 
    cs := min_src(cv);                     // current source      
    reached(cv,cs) := true;								
    if reached(cv,i) == true for all i					
        return cv;                         // all sources have reached cv
        
    min_src(cv) = argmin_{i: reached(cv,i) == false} est(cv,i);
                                           // new best source for cv
    min_est(cv) := est(v,min_src(cv))      // new best distance for cv
    										
    for each neighbor v of cv:             // update neighbour distances
        alt := d(cv,cs) + w(v,cv)          // alternate distance to cs 
        if alt < d(v,cs):                               
            d(v,cs) := alt;                // distance improved 
            est(v,cs) := d(v,cs) + h(v);   // new heuristic distance
            if est(v,cs) < min_est(v)					
                min_est(v) := est(v,cs);   // improved best distance
                min_src(v) := cs;							
\end{verbatim}

\hrule
\end{minipage}
\caption{The multisource A* algorithm. $\tt{d(u,v)}$ is the edge-weighted shortest-path distance from $u$ to $v$. The heuristic costs $\tt{h(v)}$ encourages certain vertices to be found before others. For efficiency, the ``{\tt min\_est}'' array, which maintains, for each node v, the minimum estimated distance to any unreached source, should be implemented using a priority queue.}
\label{fig:multiastar}
\end{center}
\end{figure}

The success probability of minor-embedding algorithm using multisource A* is sometimes poorer than when using Dijkstra. The reason is that each time we choose a new vertex-model to represent a vertex, we would like to minimize the overlap with other vertex-models. The multisource A* approximates this by minimizing the maximum shortest-path distance, while Dijkstra approximates this by minimizing the sum of the shortest-path distances. Thus Dijkstra is a better approximation, as it considers all overlaps rather than just the worst case. 

\begin{figure}[ht]
\begin{center}
\setlength{\unitlength}{1cm}
\begin{picture}(14.5,6)
\put(0.25,0){\includegraphics[scale = 0.35]{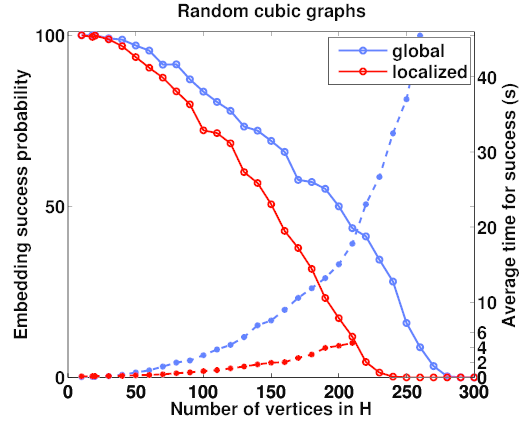}}
\put(0,5){a)}
\put(7.75,0){\includegraphics[scale = 0.35]{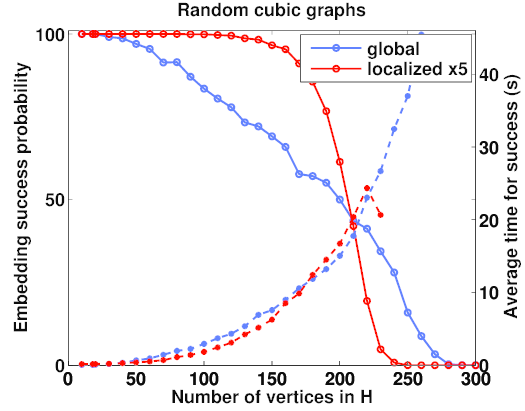}}
\put(7.5,5){b)}
\end{picture}
\end{center}
\caption{Comparing the performance of global and localized versions of the the heuristic minor-embedding algorithm. Random cubic graphs were minor-embedded into C16, which is a $2048$-vertex version of Chimera graph consisting of a $16\times 16$ grid of $K_{4,4}$'s rather than the $8 \times 8$ grid shown in Figure \ref{fig:chimera}. Each data point represents the median case of $10$ graphs of that size. Solid lines indicate success rates; dashed lines indicate running times. (a) Running the algorithm once, the localized version is always both faster and less successful than the global version. (b) Repeating the localized algorithm $5$ times on each problem instance, we obtain run times similar to the global algorithm. We see that the localized algorithm is more efficient on easier problems and less efficient on harder problems.}
\label{fig:tradeoff}
\end{figure}

However, the A* algorithm is considerably faster than the Dijkstra algorithm. Whether or not there is a benefit to this speed/performance trade-off seems to depend on the particular problem instance (see Figure \ref{fig:tradeoff} for an example).

\section{Summary and future work}
\label{sec:conclusion}

In this paper we have presented a algorithm for finding a graph $H$ as a minor of graph $G$, when both $H$ and $G$ are part of the input. To our knowledge this is the first heuristic algorithm for this problem. The algorithm has been implemented at D-Wave Systems and has proven to be effective in finding minors when $G$ and $H$ are sparse graphs with hundreds of vertices. Finding better algorithms for this problem leads directly to better use of D-Wave's quantum annealing in solving quadratic pseudo-boolean optimization problems. 

One nice feature of the algorithm presented is that even in the event of its failure, the algorithm's computations may still be useful. Define a $G$-{{\em decomposition} of $H$ to be a function $\phi:V(H) \rightarrow 2^{V(G)}$ such that the subgraph induced by each $\phi(x)$ is connected, and if $x$ and $y$ are adjacent in $H$, then there is at least one edge between $\phi(x)$ and $\phi(y)$ in $G$. A $G$-decomposition of $H$ differs from an $H$-minor of $G$ in that multiple vertices of $H$ may be represented at the same vertex of $G$. In the event of failure, the result of the minor-embedding algorithm is a $G$-decomposition. A $G$-decomposition is a loose generalization of a tree-decomposition, and algorithms for solving Ising models may take advantage of its structure. 

It seems likely that the algorithm presented here can be improved significantly by making better initial choices for the vertex models in the first iteration. Finding a good heuristic for initial placement of vertex-models is a topic for future work.

\section{Acknowledgements}
The authors thank many colleagues at D-Wave for their helpful discussions of this work, and in particular Evgeny Andriyash, Andrew King, and Jack Raymond for their comments on this paper.

\bibliographystyle{plain}
\bibliography{embedding}

\end{document}